\documentclass[pra,aps,showpacs,twocolumn]{revtex4-1}
\usepackage{dcolumn}
\usepackage{bbm}
\usepackage{amsmath,amssymb,amsfonts,latexsym,fancyhdr,graphicx,epstopdf,times,xcolor}
\newcommand{\bra}[1]{\langle#1|}
\newcommand{\ket}[1]{|#1\rangle}
\newcommand{\proj}[1]{|#1\rangle\langle#1|}
\DeclareMathOperator{\Tr}{Tr}
\DeclareMathOperator{\Id}{\mathbbm{1}}
\DeclareMathOperator{\W}{\mathcal{W}}
\graphicspath{{images/}}
\begin{document}
\title{Witnessing entanglement in hybrid systems}
%\title{Hybrid entanglement detection in trapped-ion systems}
\author{Massimo Borrelli$^1$, Matteo Rossi$^{2}$, Chiara Macchiavello$^2$, and Sabrina Maniscalco$^{1,3}$}
\affiliation{$^1$ CM-DTC, School of Engineering \& Physical Sciences, Heriot-Watt University, Edinburgh EH14 4AS, United Kingdom\\
$^2$ Dipartimento di Fisica and INFN-Sezione di Pavia, Via Bassi 6, 27100 Pavia, Italy\\
$^3$ Turku Centre for Quantum Physics, Department of Physics and Astronomy, University of
Turku, FI-20014 Turun yliopisto, Finland\\}

\begin{abstract}
We extend the definition of entanglement witnesses based on structure factors to the case in which the position of the scatterers is quantized. This allows us to study entanglement detection in hybrid systems. We provide several examples that show how these extra degrees of freedom affect the detection of entanglement by directly contributing to the measurement statistics. 
We specialize the proposed witness operators for a chain of trapped ions. Within this framework, we show how the collective vibronic state of the chain can act as an undesired quantum environment and how ions quantum motion can affect the entanglement detection. Finally, we investigate some specific cases where the method proposed leads to detection of hybrid entanglement.

\end{abstract}
%\pacs{03.65.Ta,03.65.Yz,03.67.Mn}

\maketitle

\section{Introduction}

Entanglement is a fundamental feature of quantum composite systems that has no counterpart in classical physics \cite{H3}. 
It is usually regarded as a purely quantum resource, as it is at the basis of several quantum information protocols \cite{nc}.
It is thus of great importance to reveal the presence of entanglement in an experimental set-up, as it would serve as a strong proof of its usefulness in the realization of specific quantum protocols.
To this end, the concept of entanglement witnessing was formulated  \cite{horo-ter} and proved to be an efficient tool both theoretically \cite{pra} and experimentally \cite{multi}. 
The great advantage to reveal an entangled state via witness operators is that they require fewer experimental resources, such as local measurements, than quantum state tomography .

%A possible way to find such operators for spin systems was recently proposed in Ref. \cite{chiara} and firstly verified in Ref. \cite{roma}. There, entanglement witnesses based on structure factors were introduced, and a method related to scattering experiments was explicitly proposed (see also Ref. \cite{chiaragiovanna}). Subsequent works confirmed the goodness of the method both theoretically \cite{plenio} and theoretically \cite{plenio_exp}.
A method to detect multipartite entanglement by means of structure factors was recently proposed \cite{chiara} and successfully tested in a quantum optical experiment \cite{roma}. The method was later applied to detect entanglement in a spin chain using photon scattering \cite{chiaragiovanna}. Related work in condensed matter systems was reported in \cite{plenio}. In this paper we further investigate the properties of the entanglement witnesses defined in \cite{chiara} by removing the constraint of having classically localized particles. This kind of scenario can be easily implemented in ion traps, where the ions can be cooled down so that the quantum nature of their harmonic motion emerges. Both the vibrational and the electronic degrees of freedoms of these systems can be very precisely manipulated by means of laser beams, and entangled states of the internal  (spin), external (phonons), and hybrid spin-phonons degrees of freedom have been experimentally realized. \cite{cirac,wineland}.
In this paper we investigate two related scenarios. In the first scenario we focus, as in Refs. \cite{chiara,roma,chiaragiovanna,plenio}, on spin entanglement witnesses and we ask ourselves if and how the presence of other quantum degrees of freedom (the vibrational quanta in the trapped ion case) affects the detection of entanglement between spins. In the second scenario we identify some cases for which the generalised witnesses here introduced allow to detect hybrid entanglement. This second question is of particular interest because, to the best
of our knowledge,  hybrid entanglement witnesses have not yet been
considered in the literature. When one of the system to be entangled
is, e.g., a quantum harmonic oscillator, however, revealing 
entanglement via state tomography of the total spin-harmonic oscillator system is extremely
complicated because the Hilbert space of the total system is unbounded. Hence entanglement witnesses are particularly important in this case. %This second question is of particular interest because, to the best of our knowledge,  hybrid entanglement witnesses have not yet been considered in the literature. When one of the system to be entangled is, e.g., a quantum harmonic oscillator, however, detecting entanglement of the total spin-harmonic oscillator system is extremely complicated because the Hilbert space of the total system is unbounded making tomography very challenging.

The paper is organized as follows. In Sec. \ref{sec:idea} we recall the idea of entanglement witness based on structure factors and generalize it to include quantized spatial coordinates. In Sec. \ref{sec:gauss} we present some simple examples to illustrate the behavior of the generalized witness. We then specify our analysis to  linear chains of trapped ions in Sec. \ref{sec:bosonic}, where we discuss how the collective vibrational state of the ions can influence the spin-spin entanglement detection.  In Sec. \ref{sec:hybrid} we show how the witness we introduce is able to detect hybrid entanglement, and we summarise the main results in Sec. \ref{sec:conc}.

\section{The extended entanglement witness}\label{sec:idea}

An entanglement witness is defined as an Hermitian operator $W$ that detects an entangled state $\rho$ if it has a negative expectation value for this state, $\Tr[W\rho]< 0$, while $\Tr[W\sigma]\geq 0$ for all separable states $\sigma$ \cite{horo-ter}. In this paper we will follow the construction proposed in Ref. \cite{chiara}, where a class of entanglement witnesses was introduced based on two-point spin correlation functions. These functions also define the so-called structure factors of an ensemble of $N$ particles via the expectation value of the following operator
\begin{equation}
S^{\alpha\beta}(q)=\sum_{i<j}e^{iq(r_{j}-r_{i})}S^{\alpha}_{i} S^{\beta}_{j},
\label{spinspinsf}
\end{equation}
where $S^{\alpha}_{i}$ is the $\alpha$ component of the $i^\textrm{th}$ particle spin operator and $r_{i}$ its position, which is assumed to be fixed and therefore perfectly known. 
If we focus on the case of spin $\frac{1}{2}$ particles, where ${S}^\alpha$ corresponds to the Pauli operator ${\sigma}^\alpha$, the witness operator for a general $N$-spin system is then defined as follows
\begin{equation}\label{witness1}
W(q)=\Id-\Sigma(q),
\end{equation}
where $\Id$ is the identity operator and
\begin{equation}
\Sigma(q)=\frac{1}{2}[\bar\Sigma(q) + \bar\Sigma(-q)],
\label{sigma1}
\end{equation}
with
\begin{equation}
\bar\Sigma(q)= \frac{1}{B(N,2)}
[c_{x}{S}^{xx}(q)+c_{y}{S}^{yy}(q)+c_{z}{S}^{zz}(q)].
\label{sigma2}
\end{equation}
Here $B(N,2)$ is the standard binomial factor and $c_{\alpha}\in\mathbb{R}, |c_{\alpha}|\leq 1$ for $\alpha=x,y,z$. Hence, by means of scattering measurements, it is possible to detect (multipartite) entangled states of many-particle systems.

It is worth stressing that in Eq. \eqref{witness1} one assumes completely deterministic knowledge of the scatterer's positions. For each constituent the motional degree of freedom is treated as a classical variable and, as such, it does not affect the measurement statistics. However, in some systems, such as ion traps, entanglement between the fictitious spin $\frac{1}{2}$ of each ion and the collective vibronic state can be routinely generated \cite{home}.
Hence, in the following we assume the positions of the scatterers to be operators instead of classical variables. %In order to avoid technical difficulties related to indistinguishability of the particles, we will also assume that each party is harmonically oscillating around its own equilibrium position, namely
%\begin{equation}
%r_{i}\to \hat{r}_{i}=r_{i}^{eq}+\hat{\delta r}_{i}.
%\label{ho}
%\end{equation}  
%Furthermore, in a linear chain we can assume $r_{i}^{eq}=i a$ where $a$ is the equilibrium inter-particle distance and $i=1,2,\dots,N$. 
%It is crucial at this point to underline the reason behind this quantization. 
In order to generalize the entanglement witness of Eq. \eqref{witness1} to the case in which the positions of the scatterers are quantized, let us write the spin density operator ${S}^\alpha(x)$ along the direction $\alpha$ of a $N$-particle system as \cite{chaikin}
\begin{equation}
{S}^\alpha(x)=\sum_{j=1}^{N}{S}^\alpha_{j}\otimes\delta(x-\hat{x}_{j}).
\label{spinden}
\end{equation}
In Eq. \eqref{spinden} the Dirac $\delta(\cdot)$ is meant to be an operator in the $x$-representation , and $\hat{x}_j$ is the position operator of the $j^{\textrm{th}}$ ion. (for the sake of simplicity we consider the 1-d case here). It is then possible to consider the following quantity by a suitable Fourier transform:
\begin{align}
{S}^\alpha(q)&{S}^\beta(-q)\nonumber\\
	=&\int dx_{1}dx_{2}e^{iq(x_{1}-x_{2})}{S}^\alpha(x_{1}){S}^\beta(x_{2}) \nonumber\\
=&\sum_{i<j}\int dx_{1}dx_{2}e^{iq(x_{1}-x_{2})}\delta_{1}(x_{1}-\hat{x}_{i})\delta_{1}(x_{2}-\hat{x}_{j}){S}^\alpha_{i}{S}^\beta_{j}\nonumber\\
=&\sum_{i<j}{S}^\alpha_{i}{S}^\beta_{j}e^{iq(\hat{x}_{i}-\hat{x}_{j})}.
\end{align}
%&=\sum_{i<j}\hat{S}_{i}\hat{S}_{j}\int dx_{1}e^{iqx_{1}}\delta_{1}(x_{1}-\hat{x}_{i}) \int dx_{2}e^{-iqx_{2}}\delta_{1}(x_{2}-\hat{x}_{j})=\\
As it is clear from the previous steps, this formula defines the structure factor ${S}^{\alpha\beta}(q)$ of Eq. \eqref{spinspinsf}, with the crucial difference that now the positions are regarded as intrinsically  quantum. 
The generalized entanglement witness $W(q)$ is defined in the same way as in Eqs. \eqref{witness1}-\eqref{sigma2}, but it is now a function of both spin and position operators since  each term ${S}^{\alpha\beta}(q)$ is now given by
\begin{equation}\label{quantumss}
{S}^{\alpha\beta}(q)=\sum_{i<j}{S}^\alpha_{i}{S}^\beta_{j}e^{iq(\hat{x}_{i}-\hat{x}_{j})}.
\end{equation}
It is straightforward to show that this new definition still meets the criteria for an entanglement witness \cite{chiara}. Actually, for any state of the composite system spin-position of the form $\sigma^N\otimes\rho$, being $\sigma^N=\sigma_1\otimes\sigma_2\otimes\dots\otimes\sigma_N$ and $\rho$ the spatial state of the system, we can see that
\begin{align}
&\big|\langle \Sigma(q) \rangle_{\sigma^N\otimes\rho}\big|\nonumber \\
&=\frac{1}{B(N,2)}\Big|\sum_{i<j}\Big(\sum_{\alpha=x,y,z} c_\alpha \langle S_i^\alpha S_j^\alpha\rangle_{\sigma^N}\Big)\langle\cos[q(\hat{x}_i-\hat{x}_j)]\rangle_\rho\Big|\nonumber \\
&\leq \frac{1}{B(N,2)}\sum_{i<j}\Big|\sum_{\alpha=x,y,z} c_\alpha \langle S_i^\alpha S_j^\alpha\rangle_{\sigma^N}\Big|\leq 1
\end{align}
The first inequality above comes from the fact that, whenever we deal with a composite state of the form $\sigma^N\otimes \rho$, the bound $|\langle\cos[q(\hat{x}_i-\hat{x}_j)]\rangle_\rho|\leq 1$ holds for any $\rho$. Hence, the considered witness turns out to rule out states of the form $\sigma_1\otimes\sigma_2\otimes\dots\otimes\sigma_N\otimes\rho$, namely, states fully separable in the spin and biseparable with respect to the cut spin-position. As consequence, it follows that such a witness can detect not only entanglement among spins, but also multipartite entanglement in the composite spin-position system. In particular, it is worth stressing that the above witness can in principle detect ``hybrid entanglement'', namely entanglement between degrees of freedom of different nature, such as spins versus positions. In the following we will provide some examples of hybrid entanglement detection.

We conclude this section with a recap. In a scattering experiment measurements are repeated many times and during the exposure time the particle positions are generally not fixed. The scattering intensity that is experimentally accessible is then an average over all the recorded outcomes: it can be a time average or, in most cases, an ensemble average. For a sufficiently large number of runs, this average will converge to quantum expectation values where all the relevant quantum degrees of freedom can contribute to the statistics. 
Hence, if we consider the entanglement witness in Eq. \eqref{witness1}, it is clear that whenever the scatterer's positions are to be treated quantum mechanically, as in the case of cold trapped ions, they will unavoidably affect the measurement statistics. 
%In this spirit we can thus look at them as an effective quantum environment that affects the spin-spin scattering measurements and, consequently, the ability to detect entanglement in the spin DOFs. This second scenario will be studied in Sec. \ref{sec:bosonic}.
%The expectation value for states which are fully separable in all the degrees of freedom is always non-negative. However, this is also true for states which are separable in the spin degree of freedom only. The question arises concerning what kind of entanglement or correlations we can detect. 
%
%A few examples in the following might answer this question.
\section{Spins in a harmonic potential}\label{sec:gauss}

\begin{figure}[t!]
\centering
\includegraphics[width=\columnwidth]{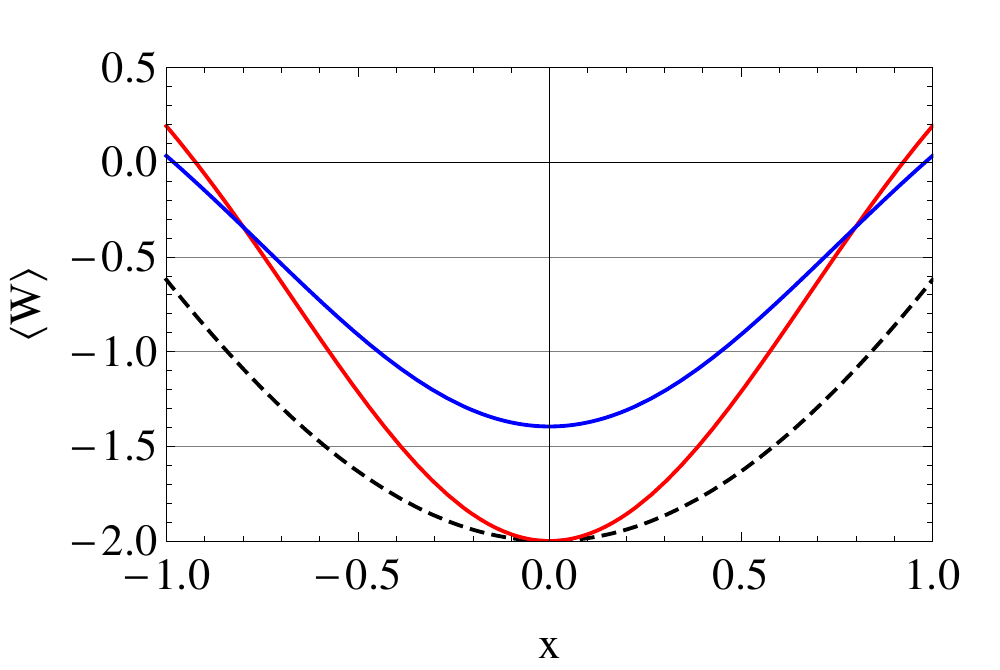}
\includegraphics[width=\columnwidth]{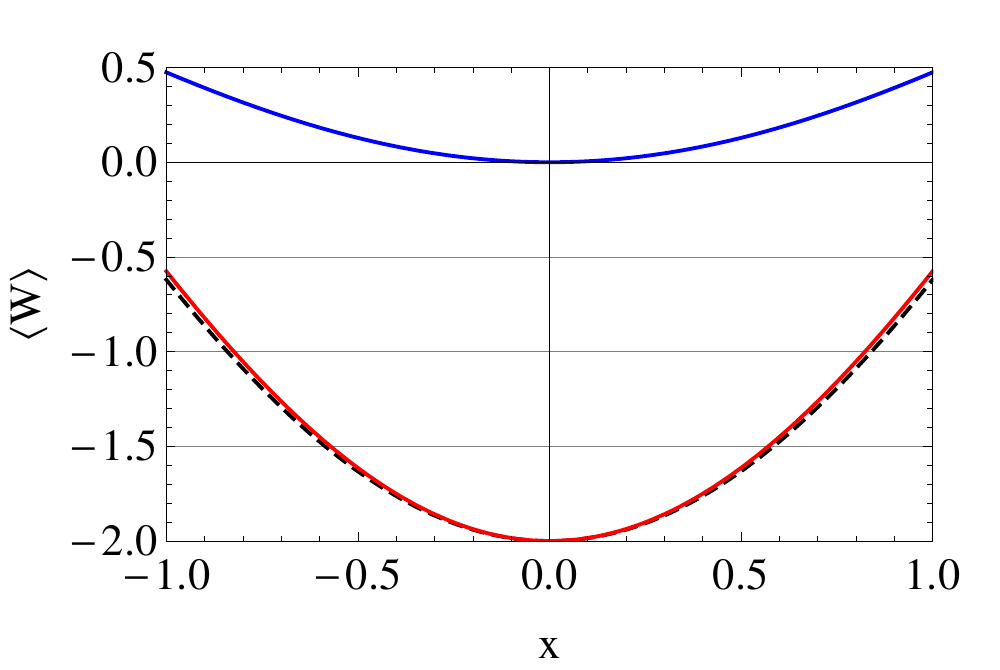}
 \caption{(Color online) Top panel: Expectation value of ${W}$ as a function of $x$ for $y=1.2$. Legend: $\langle{W}\rangle_{1}$ dashed black, $\langle{W}\rangle_{2}$ red, $\langle{W}\rangle_{3}$ blue. Bottom panel: limit $y\gg 1$.}
\label{ysm}
\end{figure}

Let us begin by considering a simple system consisting of two spin-$frac{1}{2}$ particles trapped in a double-well harmonic potential with minima centered in $x_{A}, x_{B}$. 
%whose relevant degrees of freedom are described by two spins 1/2 and 
  %e.g.  the internal degrees of freedom of two ions in a linear trap, and a quantum harmonic oscillator, e.g. the ion's collective vibrational motion. 
Each particle is described by a state of the form $\ket{\uparrow}\ket{f_A}$,
where the first ket refers to the spin state (in this case e.g. the state $\ket{\uparrow}=\sigma^z\ket{\uparrow}$) and the second encapsulates the energy contribution. For the sake of simplicity, the latter will be represented by the ground state of each harmonic oscillator, that is a Gaussian wavefunction centered in $x_{A(B)}$. %Notice that such a situation models perfectly what experimentalists often face in real set-ups, like e.g. in trapped-ion systems \textbf{giusto?![CITE]}.
We now aim at studying both how the extra continuous degrees of freedom affect the detection of the qubit-qubit entanglement, and whether the generalized witness can detect hybrid entanglement, {\it i.e.} entanglement between spins and quantum harmonic oscillators. In order to do so, let us consider the following states:
\begin{equation}
\begin{aligned}
\ket{\psi_1}&=\frac{1}{\sqrt 2}(\ket{\uparrow,\downarrow}-\ket{\downarrow,\uparrow }),\\
\ket{\psi_2}&=\frac{1}{\sqrt 2}(\ket{\uparrow,\downarrow}-\ket{\downarrow,\uparrow }) \otimes \ket{f_{A},f_{B}},\\
\ket{\psi_3}&=\frac{1}{\sqrt 2}(\ket{\uparrow,\downarrow, f_A, f_B}-\ket{\downarrow,\uparrow, f_B, f_A}),
\label{states} 
\end{aligned}
\end{equation}
where $\ket{f_J}=\sqrt{f_J(x)}\ket{x}$ with $f_J(x)= e^{-(x-x_{J})^{2}/2\sigma^{2}}/\sqrt{2\pi\sigma^{2}}$, and $J=A,B$. The above states are representative of several situations we can encounter. In state $\ket{\psi_1}$ the particles are classically localized at a distance $x_{A}-x_{B}\equiv r$. Thus, it will serve as a reference state for a comparison with states $\ket{\psi_2}$ and $\ket{\psi_{3}}$. For state $\ket{\psi_2}$ the spatial part of the wavefunction is separable from the spin state, and thus we expect just a modulation of the expectation value of the witness. Finally, for state $\ket{\psi_3}$ all the degrees of freedom are involved in a non-trivial way. The expectation values of Eq. \eqref{witness1} in the three exemplary states of Eq. (9) are 
\begin{equation}
\begin{aligned}
\langle{W}\rangle_{1}&=1+(c_{x}+c_{y}+c_{z})\cos x,\\
\langle{W}\rangle_{2}&=1+e^{-\frac{x^{2}}{y^{2}}}(c_{x}+c_{y}+c_{z})\cos x,\\
\langle{W}\rangle_{3}&=1+e^{-\frac{x^{2}}{y^{2}}} \big[(\text{c}_{x}+\text{c}_{y}) e^{-\frac{y^2}{4}}+\text{c}_{z}\cos x\big],
\end{aligned}
\label{wex1}
\end{equation}
where $x\equiv qr$ is the rescaled scattered momentum and $y\equiv r/\sigma$ quantifies the spatial overlap between the states $\ket{f_A}$ and $\ket{f_B}$. Notice that the optimal parameter choice to minimize the witness operator is $c_x=c_y=c_z=-1$ for all the states of Eq. \eqref{states}.
Fig. \ref{ysm} (top) shows $\langle W\rangle$ as a function of the rescaled scattered momenutm $x$ for the three states above and for $y=1.2$. Notice that with this choice of $y$ there is a significant overlap between the two Gaussians and so the two particle's average positions cannot be clearly identified. The dashed black curve represents the expectation value $\langle W\rangle_1$, which is negative whenever $-1\leq x\leq 1$.
The red line represents $\langle W\rangle_2$ whereas the blue line corresponds to $\langle W\rangle_3$. 
We can clearly see that in all the three cases some entanglement is present in the system as the witness takes negative values. This can be either entanglement between the two qubits or hybrid entanglement. 
In particular, the red curve is significantly more negative than the blue one. The quantum nature of particle positions results in smearing out the reference black curve, making spin-spin entanglement detection harder.
For state $\ket{\psi_3}$ the entanglement is distributed among all the different degrees of freedom and the witness $\langle W\rangle_3$ shows a regime of negativity. 

In order to further study the behaviour of $\langle W\rangle_3$ we consider the two extreme cases $y\ll 1$ and $y\gg 1$.
Whenever $y\ll 1$ holds, then the state $\ket{\psi_3}$ approaches $\ket{\psi_2}$ with $|\bra{f_A}f_B\rangle |\approx  1$, and thus the blue curve almost overlaps with the red one. On the other side, as soon as $y\gg 1$, the red curve will converge to the dashed black one, while the blue curve will get positive for any value of $x$, preventing any entanglement detection (see Fig. \ref{ysm} (bottom)).  Notice that the two-qubit reduced state of $\ket{\psi_3}$, i.e.
\begin{align}
\sigma=\frac{1}{2}&(\proj{\uparrow,\downarrow}+ \proj{\downarrow,\uparrow})\\
&-\frac{e^{-\frac{y^2}{4}}}{2}(\ket{\uparrow,\downarrow}\bra{\downarrow,\uparrow}+			\ket{\downarrow,\uparrow}\bra{\uparrow,\downarrow}).\nonumber
\end{align}
is entangled (even though the amount of entanglement would decrease exponentially in $y$). Therefore in this case $W$ fails to reveal both the entanglement between the two qubits and hybrid entanglement (being the state $\ket{\psi_3}$ GHZ-like in the limit $|\bra{f_A}f_B\rangle |\approx 0$). 
%However, we will see in Sec. \ref{sec:hybrid} that there exist witness operators able to detect hybrid entanglement too.

%\begin{figure}[h!]
%\centering
%\includegraphics[scale=0.7]{plot2.eps}
%\caption{Same as in Fig.\ref{ysm} but in the limit $y\gg 1$, here $y=15$.}
%\label{ybig}
%\end{figure}

\section{Entanglement witnessing in TRAPPED-ION SYSTEMS}\label{sec:bosonic}

\begin{figure}[t!]
\centering
\includegraphics[width=\columnwidth]{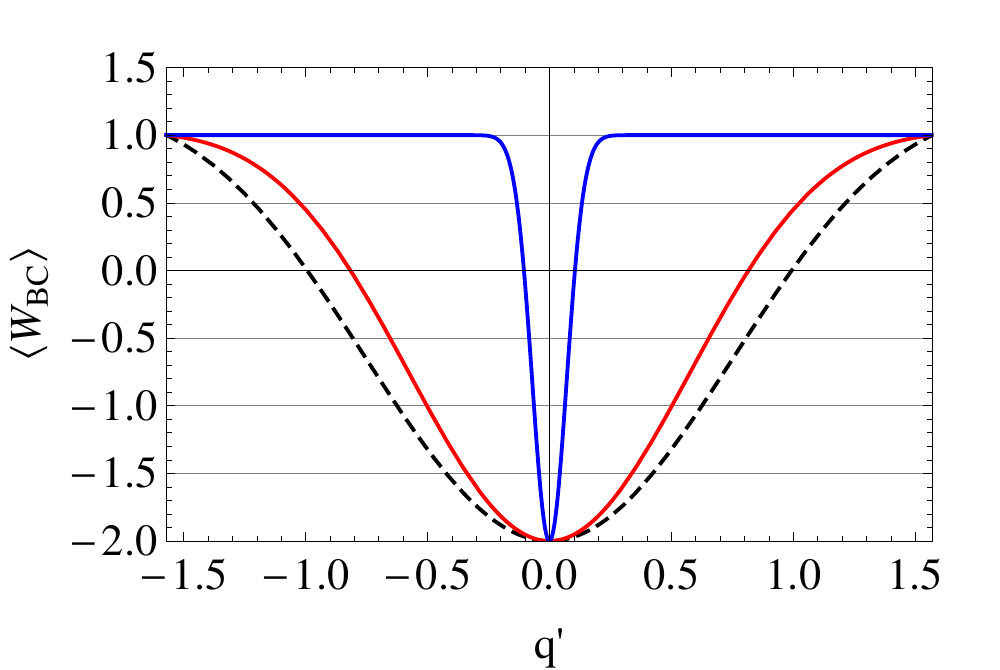}
 \caption{(Color online) Expectation value of ${W}_{BC}$ as a function of $q'$ for $\eta=1$ and $\Delta=100,1,0.01$ (dashed black, red, blue, respectively).}
\label{termicofig}
\end{figure}

In this section we will derive the specific form of the generalized witness $W$ of Eq. \eqref{witness1}, defined via Eq. \eqref{quantumss}, for trapped-ion systems. This is of great interest for basic studies of entanglement since it can be implemented with trapped-ion strings as well as Wigner crystals \cite{wigner}. 
For the sake of simplicity, we will restrict our attention to a one-dimensional system. Nevertheless this approach can be easily generalized to three dimensions.
Let us thus imagine that we have a string of $N$ ions of mass $m$, harmonically confined in a Paul trap and interacting via Coulomb repulsion. In this stable spatial configuration the ions fluctuate around their equilibrium positions. We assume the ion-ion equilibrium distance to be constant, namely $a$. This corresponds
to considering the central segment of a linear Coulomb crystal \cite{gabriele1}. For sufficiently low temperature, such a system can be approximated by a chain of interacting quantum harmonic oscillators whose equilibrium positions can be analytically ($N\le3$) and numerically computed ($N>3$) \cite{james}.
Furthermore, we can map this interacting system into a non-interacting one by standard normal-mode transformation.  If $\hat{x}_{n}$ is the position operator of the $n$-th particle, we assume that
\begin{equation}
\hat{x}_{n}\approx an+\hat{\delta} x_{n},
\label{diff}
\end{equation}
where $an$ is the equilibrium position of the $n$-th particle. 
The fluctuation operator for each ion can be expanded in terms of the ladder operators of the normal modes of the chain \cite{gabriele2}
\begin{equation}
\hat{\delta} x_{n} =\sum_{k>0}\sum_{\mu=\pm}\sqrt{\frac{\hbar}{Nm\omega_{k}}}R_{(n,k,\mu)}(\hat{a}_{k,\mu}+\hat{a}^{\dagger}_{k,\mu})
\label{xn}
\end{equation}
where $\hat{a}_{k,\mu}$ ($\hat{a}_{k,\mu}^{\dagger}$) is the annihilation (creation) for the $k$-th normal-mode with parity $\mu=\pm$ at frequency $\omega_{k}$ and the  real coefficients $R_{(n,k,\mu)}$ are the elements of the normal-mode transformation matrix. If we plug Eq. \eqref{xn} into \eqref{diff} and expand the exponential factor in Eq. \eqref{quantumss} we obtain the following result
\begin{align}
\exp\left[iq\left(\hat{x}_{n}-\hat{x}_{m}\right)\right]&= e^{iq'\left(n-m\right)}\bigotimes_{k,\mu}{D}\left[iq'\phi_{(k,\mu)}(n,m)\right]\nonumber \\
&\equiv e^{iq'\left(n-m\right)}{D}[iq'\vec{\phi}(n,m)]
\label{link}
\end{align}
where ${D}[\alpha]=\exp[\alpha \hat{a}^\dagger - \alpha^*\hat{a}]$ is the displacement operator, $q'=qa$, and we have defined
\begin{equation}
\phi_{(k,\mu)}(n,m)\equiv \sqrt{\frac{\hbar}{Nm\omega_{k}}} (R_{(n,k,\mu)}-R_{(m,k,\mu)})
\label{phi}
\end{equation}
and $\bigotimes_{k,\mu}{D}[iq'\phi_{(k,\mu)}(n,m)]\equiv{D}[iq'\vec{\phi}(n,m)]$. 
Thus, we can re express  the structure factor in Eq. \eqref{quantumss} as follows
\begin{equation}
{S}^{\alpha\beta}(q')=\sum_{n<m}{S}^{\alpha}_{n}{S}^{\beta}_{m}e^{iq'\left(n-m\right)}{D}[iq'\vec{\phi}(n,m)],
\label{sab}
\end{equation}
and the witness is then given by
\begin{equation}\label{W_BC}
{W}_{BC}(q')=\Id-\frac{1}{2}[{\Sigma}_{BC}(q')+{\Sigma}_{BC}(-q')],
\end{equation}
where
\begin{align}\label{sigma}
{\Sigma}_{BC}(q')=1-\frac{1}{B(N,2)}\sum_{n<m}&\sum_{\alpha=x,y,z}c_{\alpha}{\sigma}_{n}^{\alpha}{\sigma}_{m}^{\alpha}\\
&\times e^{iq'\left(n-m\right)}{D}[iq'\vec{\phi}(n,m)].\nonumber
\end{align}
%We stress that in trapped-ion systems the spin DOF is usually encoded in two electronic states of the ion.  
Whenever the state of the composite system is $\sigma\otimes \rho$, with $\sigma$ and $\rho$ being the states of the internal degrees of freedom of the ion chain and the external (vibrational) deegres of freedom, respectively, the expectation value of the witness \eqref{W_BC} reads as follows
\begin{align}\label{W_charc}
\langle {W}_{BC}(q')\rangle=1-\frac{1}{B(N,2)}&\sum_{n<m}\sum_{\alpha=x,y,z}c_{\alpha}\langle{\sigma}_{n}^{\alpha}{\sigma}_{m}^{\alpha}\rangle_{\sigma} \\
\times \text{Re}&\{ e^{iq'\left(n-m\right)}\langle{D}[iq'\vec{\phi}(n,m)]\rangle_{\rho}\}.\nonumber
\end{align}
Therefore, in this case we can, in principle, relate the expectation value of the entanglement witness to the characteristic function of the vibrational state,  
$C_{\W}(\alpha)=\langle {D}(\alpha)\rangle_\rho$. The quantity $ \text{Re}\{\dots\}$ appearing in Eq. \eqref{W_charc} can be recast as
\begin{align}
\text{Re}\{\dots\}=\cos &[q'(n-m)]\text{Re}\{C_{\W}(iq'\vec{\phi}(n,m))\}\\
			&-\sin[q'(n-m)]\text{Im}\{C_{\W}(iq'\vec{\phi}(n,m))\}.\nonumber
\end{align}
Recall that the characteristic function $C_{\W}(\alpha)$ is related to the Wigner representation $\W(\alpha)$ via the inverse Fourier transform \cite{sir}.
As a straightforward consequence we see that, whenever the collective state of the spins and modes is factorized, the witness $W_{BC}$ is just modulated according to the characteristic function of the phononic bath, as sampled in some specific points of the phase space. Hence, in this scenario the vibrational degrees of freedom simply act as non-classical noise affecting the scattering readout and the spin-spin entanglement detection. As a very simple but meaningful example, let us consider the situation where we deal with two particles and a single-mode bath. Then, if we assume the composite system to be in the state
\begin{equation}
\rho=\frac{1}{2}(\ket{\uparrow,\downarrow}+\ket{\downarrow,\uparrow})
		(\bra{\uparrow,\downarrow}+\bra{\downarrow,\uparrow})
					\otimes \rho_{\pi/a,T},
\label{termico}
\end{equation}
where $\rho_{\pi/a,T}$ is a thermal state at temperature $T$ for the mode at $k=\pi/a$. It is possible to study the limit to the entanglement detection caused by the temperature of the system or, equivalently, by the fact that the first collective vibrational mode of the chain is in a thermal state. In fact, the expectation value of $W_{BC}(q')$ turns out to be 
\begin{equation}
\langle W_{BC}(q') \rangle=1-e^{-\frac{1}{2}q'^{2}\eta^{2}\coth\frac{\Delta}{2}}(c_{x}+c_{y}-c_{z})\cos q' ,
\label{wt}
\end{equation}
where $\eta=a^{-1/4}\sqrt{\hbar/Q\sqrt{8m}}$ with $Q$ being the atom's electric charge, $\Delta=\hbar\omega_{\pi/a}/k_{B}T$, and $\omega_{\pi/a}$ is the frequency
of the $k=\pi/a$ mode. The expectation value of the witness in this case is shown in Fig. \ref{termicofig} as a function of the rescaled momentum $q'$ for different values of the energy-temperature ratio. In this plot the values of the parameters are the same as in the experiment reported in \cite{birkl,waki}, corresponding to $a\approx 33 \mu m$ for $^{24}\textrm{Mg}^{+}$ atoms. Fig. \ref{termicofig} clearly shows that increasing the temperature will result in drastically limiting the entanglement detection of the spin state as the range of $q'$ momentum corresponding to $\langle W_{BC}(q') \rangle<0$ shrinks. 

%\textbf{[stato squeezed?!]}
%Within this framework the influence of highly non-classical states of the single-mode bath can be addressed too. This is the case for instance if we consider the boson in a squeezed state of the form...\textbf{to do: fare il calcolo e vedere che succede, poi commenta. la funzione caratteristica si trova anche in \cite{un} ma e' un po' complicata, ne conosci di gia' semplificate?}.

\section{Detection of Hybrid Entanglement}\label{sec:hybrid}

In this section we shall study the behaviour of the witness ${W}_{BC}$ for some relevant quantum states that can be generated in trapped-ion systems via a series of suitable laser pulses \cite{home,blattwineland}. For the sake of simplicity, we shall restrict our analysis to a system consisting of two ions whose internal degrees of freedom are coupled to the collective vibrational mode. We will show that even in such a simple system hybrid entanglement can be detected for some specific states, at least in principle. %The generalization to more complex scenarios is  
We denote by $\ket{n}$ the Fock state with $n$ excitations of the $k=\pi/a$ vibrational mode. Let us thus consider the following states of the composite system %\textbf{(conoscete altri stati interessanti per i trapped-ions?!)}
\begin{align}\label{states2}
\ket{\phi_1}&=\frac{1}{\sqrt 2}(\ket{\uparrow,\downarrow,\alpha}+\ket{\downarrow,\uparrow,-\alpha}),\nonumber \\
\ket{\phi_2}&=\frac{1}{\sqrt 2}(\ket{\uparrow,\downarrow,0}+\ket{\downarrow,\uparrow,1}),\\
\phi_3 &=p\proj{\Psi^+}\otimes \proj{0} + (1-p)\proj{\downarrow,\downarrow}\otimes\proj{1}, \nonumber
\end{align}
where $\ket{\Psi^+}=\frac{1}{\sqrt 2}(\ket{\uparrow,\downarrow}+\ket{\downarrow,\uparrow})$, and $\ket{\alpha}$ is the coherent state of the mode, i.e. $\ket{\alpha}=e^{-|\alpha|^2/2}\sum_n\frac{\alpha^n}{\sqrt{n!}}\ket{n}$.
%which we choose real for the sake of simplicity.
The first two represent a superposition of states in all the degrees of freedom, while the last state usually results when the excitation, at the beginning present in the spin system, spontaneously decoheres to the bath.
With the help of Eq. \eqref{W_BC}, it is not difficult to see that
\begin{align}\label{wex2}
\langle W_{BC}\rangle_{1}=1+ & e^{-\frac{1}{2}q'^2\eta^2}
	\big[c_{z}\cos[2q'\eta\alpha_{\text{Re}}]\nonumber \\
		&-(c_x+c_y)e^{-2|\alpha|^2-2q'\eta\alpha_{\text{Im}}}\big]\cos q',\nonumber\\
\langle W_{BC}\rangle_{2}=1+ & \frac{1}{2}e^{-\frac{1}{2}q'^2\eta^2}
		\big[c_{z}(2-q'^2\eta^2)\cos q'\nonumber\\ 
		&-2 (c_{x}+c_{y})q'\eta \sin q'\big],\\
\langle W_{BC}\rangle_{3}=1+&	e^{-\frac{1}{2}q'^2\eta^2}\big[c_z(p(2-q'^2\eta^2)+q'^2\eta^2-1)\nonumber \\
&-p(c_x+c_y)\big]\cos q',\nonumber
\end{align}
where $\alpha=\alpha_{\text{Re}}+i\alpha_{\text{Im}}$ and $\eta=a^{-1/4}\sqrt{\hbar/Q\sqrt{8m}}$, as in the previous section. Notice that, since these states are never factorized with respect to different degrees of freedom, then formula \eqref{W_charc} cannot be applied here. Furthermore, the optimal parameters are still $c_x=c_y=1$ and $c_z=-1$.
\begin{figure}[t!]
\centering
\includegraphics[width=\columnwidth]{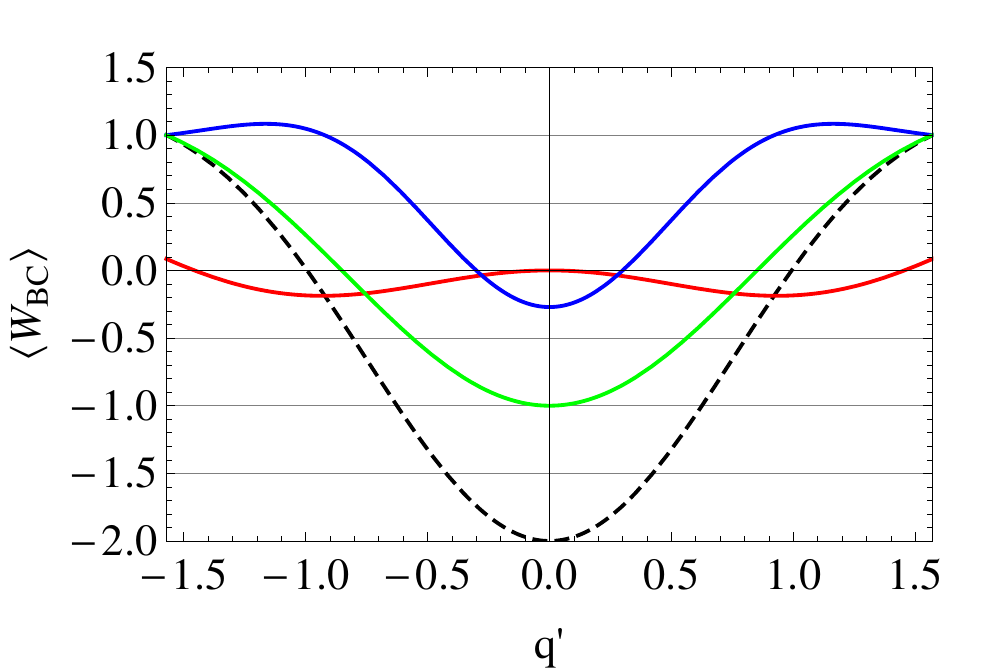}
\caption{(Color online) Expectation value of ${W}_{BC}$ in terms of $q'$ for the states in Eq. \eqref{states2}: $\langle{W}_{BC}\rangle_{1}$ (with $\alpha=0$) dashed black, $\langle{W}_{BC}\rangle_{1}$ (with $\alpha=1$) blue , $\langle{W}_{BC}\rangle_{2}$ red, $\langle{W}_{BC}\rangle_{3}$ green.}
\label{plot3}
\end{figure}
The expectation values are showed in Fig. \ref{plot3} as functions of the reduced wave-vector  $q'$, with the same choice of parameters as in Refs. \cite{birkl,waki}, reported in the previous section. The dashed black line represents $\langle W_{BC}\rangle_1$, i.e. the expectation value of $W_{BC}$ on the state $\ket{\phi_1}$ with $\alpha=0$. The state turns out to be separable with respect to different degrees of freedom, {\it i.e.}, no hybrid entanglement is present, and the presence of the extra mode results in a modulation of the expectation value according to the characteristic function of the Fock state $\ket{0}$. 
The expectation value $\langle W_{BC}\rangle_1$ for $\alpha=1$ is instead depicted by the blue line. As it is clear from the plot, entanglement is still detected, even if for a smaller range of values of $q'$. Notice that the blue curve approaches the dashed black one whenever $|\alpha|\ll 1$. Further numerical studies show that the entanglement detection is no longer guaranteed for values of $|\alpha|$ slightly larger than $1$. It is worth stressing that  $|\alpha|$ plays here exactly the same role as the parameter $y$ played in the context of the Gaussian delocalization. Hence, the comments we made for the state $\ket{\psi_3}$ of Eq. \eqref{states} still hold here.
The main feature displayed in Fig. \ref{plot3} is nevertheless the unexpected behavior of the witness $W_{BC}$ for the state $|\phi_{2}\rangle$. In this case two minima are symmetrically located with respect to $q'=0$ which, instead, appears to be the optimal scattering wave-vector for all the other states. It is worth stressing that, in this case, while the state of the total hybrid system is entangled (as shown by the witness), the reduced spin state is clearly separable. This indicates that witness does capture hybrid entanglement between the spins and the oscillatory mode. This instance definitely shows the hybrid-entanglement detection by the use of a witness with extra quantum degrees of freedom, such as e.g. an external bosonic mode.
Last but not least, the noisy state $\phi_3$, modelling the situation where the single excitation of the spins is lost into the oscillatory mode, is detected as far as $p$ is close enough to $1$. In fact, for $q'=0$ the noise threshold in order to have entanglement detection is $p\geq 1/2$. 
%Furthermore, as one would expect, the expectation value $\langle W_{BC}\rangle_3$ approaches $\langle W_{BC}\rangle_1$ for $\alpha=0$ as $p\approx 1$. 

%\section{Experimental implementation for trapped ions.}
%In \cite{chiaragiovanna} an experimental proposal to detect multi-partite entanglement in a chain of two-level atoms is presented. It relies on a four-level scheme composed by two ground states plus two excited states coupled via lasers and via a cavity mode. The output intensity of the cavity mode is the quantity to be measured and, under standard approximations, it directly links to the entanglement witness introduced in \cite{chiara} 

\section{Conclusions}\label{sec:conc}
In this work we have generalized the class of entanglement witnesses based on structure factor to the case of hybrid systems possessing both discrete and continuous variable degrees of freedom. More precisely we have looked at the situation in which the position of the scatterers is quantized. Our approach is not only a generalization to the full quantum case, but it also allows to deal with the delicate and important issue of entanglement detection in hybrid systems.

To illustrate our results we have considered a simple model where two particles were equipped with a spatially delocalized state and showed how this extra degree of freedom indeed affects the entanglement detection. Then we have explicitly developed the form of the witness operator for the case of a chain of trapped ions.
Within this framework, we have first studied a connection between the expectation value of the witness operator and the characteristic function of the vibrational state, showing how such a connection affects the spin-spin entanglement detection. 
Moreover, we have further pointed out some instances where
we could also detect some hybrid entanglement, that is entanglement between different degrees of freedom, such as spins and bosonic modes. 

The theoretical results presented here can be applied to an experimental set-up by following the method proposed in Ref. \cite{chiaragiovanna}, or in an experimental implementation such as the one considered in \cite{plenio_exp}.

\section*{Acknowledgments}
M. B. would like to acknowledge the Scottish Doctoral Training Centre in Condensed Matter Physics (EPSRC CM-DTC) for financial support.

\end{document}